\documentclass[letterpaper, 10pt]{IEEEtran}
\usepackage{cite}
\usepackage{amsmath,amssymb,amsfonts}
\usepackage{algorithmic}
\usepackage{graphicx}
\usepackage{textcomp}
\usepackage{xcolor}
\usepackage{mathrsfs}
\usepackage{psfrag}
\usepackage{cancel}
\usepackage{datetime}
\usepackage{comment}

\makeatletter
\let\NAT@parse\undefined
\makeatother
\usepackage{hyperref}
\usepackage[normalem]{ulem}

\def\mc{\mathcal}

\definecolor{mcl}{HTML}{036EF6}

\def\BibTeX{{\rm B\kern-.05em{\sc i\kern-.025em b}\kern-.08em
    T\kern-.1667em\lower.7ex\hbox{E}\kern-.125emX}}
\begin{document}
\title{\LARGE \bf Negative Imaginary Control Using Hybrid Integrator-Gain Systems: Application to MEMS Nanopositioner
}

\author{{Kanghong Shi},$\quad${Nastaran Nikooienejad},$\quad${Ian R. Petersen, \IEEEmembership{Life Fellow, IEEE}},\\ and {S. O. Reza Moheimani, \IEEEmembership{Fellow, IEEE}}
\thanks{This work was supported by the Australian Research Council under grant DP190102158, and partially by the UTD Center for Atomically Precise Fabrication of Solid-state Quantum Devices.}
\thanks{Kanghong Shi and Ian R. Petersen are with the School of Engineering, College of Engineering, Computing and Cybernetics, Australian National University, Canberra, Acton, ACT 2601, Australia. Nastaran Nikooienejad and S. O. Reza Moheimani are with the Erik Jonsson School of Engineering and Computer Science, The University of Texas at Dallas, Richardson, TX 75080 USA. Corresponding author: S. O. Reza Moheimani. {\tt kanghong.shi@anu.edu.au}, {\tt nastaran.nikooienejad@gmail.com}, {\tt ian.petersen@anu.edu.au}, {\tt reza.moheimani@utdallas.edu}}}

\newtheorem{definition}{Definition}
\newtheorem{theorem}{Theorem}
\newtheorem{conjecture}{Conjecture}
\newtheorem{lemma}{Lemma}
\newtheorem{remark}{Remark}
\newtheorem{corollary}{Corollary}
\newtheorem{assumption}{Assumption}

\maketitle
\thispagestyle{empty}
\pagestyle{empty}

\begin{abstract}
In this paper, we propose a new approach to address the control problem for negative imaginary (NI) systems by using hybrid integrator-gain systems (HIGS). We investigate the single HIGS of its original form and its two variations, including a multi-HIGS and the serial cascade of two HIGS. A single HIGS is shown to be a nonlinear negative imaginary system, and so is the multi-HIGS and the cascade of two HIGS. We show that these three types of HIGS can be used as controllers to asymptotically stabilize linear NI systems. The results of this paper are then illustrated in a real-world experiment where a 2-DOF microelectromechanical system nanopositioner is stabilized by a multi-HIGS.
\end{abstract}


\section{INTRODUCTION}
Negative imaginary (NI) systems theory was introduced in \cite{lanzon2008stability,petersen2010feedback} to provide an alternative approach to the robust control of flexible structures \cite{junkins1993introduction,preumont2018vibration,halim2001spatial,pota2002resonant}. Flexible structures usually have highly resonant dynamics, for which the traditional negative velocity feedback control may not be suitable. Also, unmodelled uncertainties in such systems can lead to poor performance or even instability if the controller is not designed to be robust against them \cite{petersen2010feedback,petersen2016negative}. As NI systems theory uses positive position control, it is efficient towards the control of flexible structures and has attracted attention since was introduced in 2008 (see \cite{cai2010stability,mabrok2014generalizing,song2012negative,xiong2010negative,bhowmickoutput,bhikkaji2011negative,wang2015robust}, etc). NI systems theory can be regarded as a complement to the positive real (PR) system theory. One limitation of PR systems theory is that it can only deal with systems having relative degree of zero and one \cite{brogliato2007dissipative}. NI systems, which can be regarded as the cascade of a PR system and an integrator, can have relative degree of zero, one and two \cite{shi2021necessary}. Typical mechanical examples of NI systems arise in systems with colocated force actuators and position sensors. Roughly speaking, a transfer function matrix $G(s)$ is said to be NI if it is stable and $j(G(j\omega)-G(j\omega)^*)\geq 0$ for all frequencies $\omega>0$. NI systems can be stabilized using strictly negative imaginary (SNI) systems. Under some assumptions, the positive feedback interconnection of an NI system $G(s)$ and an SNI system $R(s)$ is asymptotically stable if and only if the DC loop gain of the interconnection is less than unity; i.e., $\lambda_{max}(G(0)R(0))<1$ (e.g., see \cite{lanzon2017feedback}). NI systems theory has been applied in a wide variety of systems, including nanopositioning control \cite{mabrok2013spectral,das2014mimo,das2014resonant,das2015multivariable}, control of lightly damped structures \cite{cai2010stability,rahman2015design,bhikkaji2011negative}, cooperative control for networked multi-agent systems \cite{wang2015robust,skeik2019distributed,shi2023output}, etc.

NI systems theory was extended to nonlinear systems in 2018 \cite{ghallab2018extending} considering that many NI systems are of a nonlinear nature. A nonlinear system is said to be NI if it is dissipative with respect to the inner product of its input and the time derivative of its output. The papers \cite{ghallab2018extending, shi2021robust,shi2023output} showed that a nonlinear negative imaginary (NNI) system can be asymptotically stabilized by various classes of strict negative imaginary systems applied in positive feedback when certain assumptions are satisfied. This extension not only allows for nonlinear plants but also brings flexibility in the choice of controllers. Nonlinear controllers or even hybrid controllers can now be considered in the control of NI plants in order to achieve certain control objectives. One example of such a controller is the hybrid integrator-gain system (HIGS), which can improve control performance for single-input single-output (SISO) NI systems \cite{shi2022negative}.


Linear feedback control has inherent limitations which have been discussed in \cite{middleton1991trade,freudenberg2000survey}. The Bode's phase-gain relationship shows a trade-off between system performance and its robustness \cite{bode1945network,horowitz1963synthesis}. That is, the desired large low-frequency gain and small high-frequency gain can only be achieved at the cost of a lower stability margin \cite{chen2000stability}. To be specific, an integrator, which is commonly used to eliminate the steady state error \cite{li2006pid} introduces a $90^\circ$ phase lag. The time delay caused by such a phase lag will unavoidably lead to an overshoot and even instability \cite{hu1997zero,Clegg_1958}. Clegg introduced a nonlinear integrator in \cite{Clegg_1958}, known as Clegg integrator, to overcome these limitations. The output of the Clegg integrator is reset to zero whenever its input crosses zero. With such a resetting mechanism, the input and output of a Clegg integrator always have the same sign. The describing function of a Clegg integrator has a magnitude slope identical to a linear integrator but a phase lag of only $38.1^\circ$. In comparison to a linear integrator, the reduction of $51.9^\circ$ of phase lag will lead to a significant reduction in time delay and, as a consequence, the overshoot. The concept of output resetting was generalized to a first-order filter with the transfer function $1/(s+b)$ in \cite{horowitz1975non}. This is known as the first-order reset element (FORE); see \cite{chait2002horowitz,Zaccarian_FORE_2005} for summaries of reset control systems. A concrete example provided in \cite{beker2001plant} shows that a reset control system can meet control objectives that are unachievable for any linear controller: under a FORE control, the plant tracks a unit step reference with no overshoot even for a large rise time, i.e., when the bandwidth is low (see \cite{middleton1991trade}).

A drawback of conventional reset control systems such as Clegg integrators and FOREs is that they generate discontinuous control signals once the reset happens. Discontinuous control signals can cause chattering which may excite high-frequency dynamics and lead to poor performance or even instability \cite{bartolini1989chattering,khalil2002nonlinear}.
To overcome this drawback of reset control systems, the HIGS was introduced in~\cite{deenen2017hybrid}. Instead of resetting the state to zero, the HIGS alternates between integrator and gain modes resulting in a continuous (but non-smooth) control signal. This prevents the excitation of high-frequency harmonics induced by conventional reset control systems. The input-output relation of the HIGS is also restricted to a sector in which the HIGS operates as an integrator. The tendency to violate the sector constraint in the integrator mode enforces switching to the gain mode, ensuring that the input and output of the HIGS would have identical signs \cite{heertjes2019hybrid}. The frequency response of the describing function of a HIGS has the same $38.1^\circ$ of phase lag as the reset control system \cite{deenen2017hybrid}. Hence, the HIGS has similar advantages as the reset control systems described above in terms of overcoming the limitations of linear controllers. The paper \cite{Eijnden_HIGS_Overshoot_limitation_2020} illustrated the advantages of HIGS using a concrete example where the overshoot is completely eliminated by HIGS control, which is unachievable by any linear controller.

The demand for high-precision, high-speed, and reliable mechatronics systems has tremendously increased, and to address this need, the HIGS element has been exploited successfully in the semiconductor industry for motion tracking~\cite{deenen2017hybrid, Eijnden_HIGS_motion_control_2018, Gruntjens_HIGS_Lens_motion_2019}, vibration isolation and damping~\cite{heertjes2019hybrid, Achten_HIGS_Skyhook_thesis_2020, Baaij_HIGS_positive_real_systems}. In~\cite{deenen2017hybrid}, a HIGS-based $\text{PI}^2\text{D}$ controller is designed and applied to a wafer stage system of an industrial wafer scanner. Owing to the enhanced phase behaviour of HIGS-based filters compared to the linear counterparts, HIGS-based second-order low-pass filter~\cite{Eijnden_HIGS_motion_control_2018} and HIGS-based notch filter~\cite{Hebers_HIGS_Notch_2020} are proposed to achieve substantial low-frequency disturbance rejection and increase the bandwidth of a wafer scanner. A HIGS-based bandpass filter is also constructed in~\cite{heertjes2019hybrid}, featuring a series connection of two HIGS elements applied for vibration isolation. By replacing a standard integrator with the HIGS element, \cite{Eijnden_HIGS_Overshoot_limitation_2020} demonstrates a novel application of the HIGS in reducing overshoot in linear time-invariant plants having a real unstable pole. The HIGS is also studied in multivariable configuration~\cite{Achten_HIGS_Skyhook_thesis_2020} applied to a multivariable active vibration isolation system in the form of a HIGS-based bandpass filter.

Stability and performance analysis of closed-loop systems featuring HIGS is challenging due to the nonlinear nature of this hybrid system.
Based on the stability analysis of reset control systems in general~\cite{van2017frequency}, a frequency-domain approach is proposed in~\cite{deenen2017hybrid} to graphically verify the stability of a controlled system with the HIGS using the measured frequency response data of the linear part and the Kalman-Yakubovich-Popov (KYP) lemma. In this approach, the closed-loop system is rearranged in Lur'e form by isolating the nonlinearity from the linear counterpart, thus the input-to-state stability (ISS) of the closed-loop system is guaranteed based on the detectability of the HIGS element and the circle criterion. This results in frequency-domain conditions, less stringent than the strictly positive criterion~\cite{van2017frequency}. Using a modified version of the circle criterion, the stability of the closed-loop system in multiple-input multiple-output (MIMO) configuration is also investigated in~\cite{Achten_HIGS_Skyhook_thesis_2020}.

Since the switching strategy in HIGS is not taken into account in the proposed stability analysis of the controlled systems featuring HIGS, the frequency-domain conditions are a conservative estimate of the stability. This has been addressed in~\cite{Eijnden_frequency_stability_HIGS_2021} by proposing novel conditions that guarantee the existence of the Lyapunov functions in subregions of the state-space where the HIGS is active. The stability of nonlinear closed-loop systems with the HIGS element is also explored through a time-domain approach where an ISS condition is proposed in terms of linear matrix inequalities (LMIs) that guarantee the existence of a piecewise quadratic Lyapunov function~\cite{deenen2021projection}. This approach is less conservative compared to the frequency-domain approach~\cite{deenen2021projection}.

In this paper, we investigate the application of HIGS on the robust control of linear NI systems. We propose different types of variations of the original HIGS which was introduced in \cite{deenen2017hybrid}. To be specific, we investigate a single HIGS, a multi-HIGS, and the cascade of two HIGS elements. A conference version of this paper \cite{shi2022negative} showed that a HIGS is an NNI system and can be used to asymptotically stabilize a SISO NI system. The present paper aims to explore different types of HIGS-based NI controllers constructed by connecting multiple single HIGS elements either in parallel or in series. The parallel connected HIGS, called multi-HIGS, helps extend the results in \cite{shi2022negative} to MIMO systems, which is the main contribution of this paper. We provide a more intuitive description of the multi-HIGS, which was originally introduced in \cite{Achten_HIGS_Skyhook_thesis_2020}. It is shown that a multi-HIGS is also an NNI system. For any MIMO NI plant with a minimal realization, there always exists a multi-HIGS controller that asymptotically stabilizes the NI plant. As for the serial cascade of multiple HIGS, we only investigate the case of two serial cascaded HIGS because NI systems can only have relative degree less than or equal to two \cite{shi2021necessary}. It is shown that the cascade of two HIGS is also an NNI system and can be used as an alternative to the single HIGS controller in stabilizing a SISO NI system.

The NI property of the HIGS elements and the theoretical stability results motivate a methodology in the control of a certain class of systems in practice. That is, using HIGS in the control of flexible structures with colocated force actuators and position sensors. This methodology is applied in this paper to control one such system -- a microelectromechanical system (MEMS) nanopositioner. A real-world experiment was implemented, where a 2-DOF MEMS nanopositioner was controlled by a multi-HIGS controller.

The rest of the paper is organized as follows. Section \ref{sec:preliminaries} provides some preliminaries on HIGS and NI systems. Section \ref{sec:SISO HIGS} shows the NI property of a single SISO HIGS of its original form as given in \cite{deenen2017hybrid}. It is also shown in Section \ref{sec:SISO HIGS} that a single HIGS can stabilize a SISO linear NI system. Sections \ref{sec:cascaded HIGS} and \ref{sec:MIMO_HIGS} investigate two variations of HIGS; i.e., two cascaded HIGS and multi-HIGS, respectively. System models of these two variations of HIGS are given. We show their NNI properties, which are then used in stabilizing linear NI plants. In Section \ref{sec:example}, the proposed results are applied to the control of a MEMS nanopositioner. Section \ref{sec:conclusion} concludes the paper.

Notation:  $\mathbb R$ denotes the set of real numbers. $\mathbb R^{m\times n}$ denotes the space of real matrices of dimension $m\times n$. $A^T$ denotes the transpose of a matrix $A$. $A^{-T}$ denotes the transpose of the inverse of $A$; i.e., $A^{-T}=(A^{-1})^T=(A^T)^{-1}$. $\lambda_{max}(A)$ denotes the largest eigenvalue of a matrix $A$ with real spectrum. For a symmetric matrix $P$, $P>0\ (P\geq 0)$ denotes the fact that the matrix $P$ is positive definite (positive semi-definite) and $P<0\ (P\leq 0)$ denotes the fact that the matrix $P$ is negative definite (negative semi-definite). Let $\theta_i \in \mathbb R^N$ denote the standard unit vector; i.e., the $i$-th element of $\theta_i$ is one and all other elements are zeros.

\section{PRELIMINARIES}\label{sec:preliminaries}
\subsection{Hybrid Integrator-Gain Systems}
A SISO hybrid integrator-gain system (HIGS) $\mathcal{H}$ is represented by the following differential algebraic equations (DAEs)~\cite{deenen2017hybrid}:
	\begin{equation}\label{eq:HIGS}
		\mathcal{H}:
		\begin{cases}
			\dot{x}_h = \omega_h e, & \text{if}\, (e,u,\dot{e}) \in \mathcal{F}_1\\
			x_h = k_he, & \text{if}\, (e,u,\dot{e}) \in \mathcal{F}_2\\
			u = x_h,
		\end{cases}
	\end{equation}
where $x_h,e,u \in \mathbb{R}$ denote the state, input, and output of the HIGS, respectively. Here, $\dot{e}$ is the time derivative of the input $e$, which is assumed to be continuous and piecewise differentiable. Also, $\omega_h \in [0,\infty)$ and $k_h \in (0, \infty)$ represents the integrator frequency and gain value, respectively. These tunable parameters allow for desired control performance. The sets $\mathcal{F}_1$ and $\mathcal{F}_2 \in \mathbb{R}^3$ determine the HIGS modes of operation; i.e. the integrator and gain modes, respectively. The HIGS is designed to operate under the sector constraint $(e,u,\dot{e})\in \mc F$ (see \cite{deenen2017hybrid,Achten_HIGS_Skyhook_thesis_2020}) where
\begin{equation}\label{eq:F}
	\mathcal{F} = \{ (e,u,\dot{e}) \in \mathbb{R}^3 |\, eu \geq \frac{1}{k_h}u^2\},
\end{equation}
and $\mathcal{F}_1$ and $\mathcal{F}_2$ are defined as
\begin{align*}
	\mathcal{F}_1& := \mathcal{F} \setminus \mathcal{F}_2;\\
	\mathcal{F}_2& := \{(e,u,\dot{e}) \in \mathbb{R}^3 | u = k_he\quad \text{and}\quad  \omega_he^2 > k_he\dot{e}\}.
\end{align*}
HIGS (\ref{eq:HIGS}) operates in the integrator mode unless the HIGS output $u$ is on the boundary of the sector $\mathcal{F}$, and tends to exit the sector; i.e. $(e,u,\dot{e}) \in \mathcal{F}_2$. In this case, the HIGS is enforced to operate in the gain mode. At the time instants when switching happens, the state $x_h$ still remains continuous, as can be seen from (\ref{eq:HIGS}).

\subsection{Negative Imaginary Systems}
\begin{lemma}(NI Lemma)\cite{xiong2010negative}\label{lemma:NI}
Let $(A,B,C,D)$ be a minimal state-space realization of an $p\times p$ real-rational proper transfer function matrix $G(s)$ where $A\in \mathbb R^{n\times n}$, $B\in \mathbb R^{n\times p}$, $C\in \mathbb R^{p\times n}$, $D\in \mathbb R^{p\times p}$. Then $R(s)$ is NI if and only if:

1. $\det(A)\neq 0$, $D=D^T$;

2. There exists a matrix $Y=Y^T>0$, $Y\in \mathbb R^{n\times n}$ such that
\begin{equation*}
	AY+YA^T\leq 0,\qquad \textnormal{and} \qquad B+AYC^T=0.
\end{equation*}	
\end{lemma}

Considering a general nonlinear system
\begin{subequations}\label{eq:general_nonlinear_system}
\begin{align}
	\dot{x} &= f(x,u),\\
	y &= h(x),
\end{align}
\end{subequations}
where $x\in \mathbb R^n$, $u,y\in \mathbb R^p$ are the state, input and output of the system, respectively.
\begin{definition}(NNI Systems)\cite{ghallab2018extending,shi2021robust}\label{def:nonlinear_NI}
	A system of the form (\ref{eq:general_nonlinear_system}) is said to be an NNI system if there exists a positive definite continuously differentiable storage function $V:\mathbb{R}^n \to \mathbb{R}$ such that
	\begin{equation*}
		\dot{V}(x(t)) \leq u(t)^T\dot{y}(t), \quad \forall\, t \geq 0.
	\end{equation*}
\end{definition}

\section{A SINGLE HIGS}\label{sec:SISO HIGS}
In this section, we present that a single HIGS, represented by (\ref{eq:HIGS}),  is NNI. Then, we show that a single HIGS can be used as a controller for a SISO NI system by proving that the positive feedback interconnection of an NI system and a HIGS is asymptotically stable.
\subsection{NNI Property of SISO HIGS}
We first present a property of the HIGS (\ref{eq:HIGS}) in Lemma \ref{lemma:F implication}, which is implied by the sector constraint (\ref{eq:F}). This property will be used later in the proofs of the main results.
\begin{lemma}\label{lemma:F implication}
Consider a HIGS element with the system model (\ref{eq:HIGS}). This system satisfies
\begin{equation*}
ex_h-k_he^2\leq 0,	
\end{equation*}
where the equality only holds when $x_h = k_he$.
\end{lemma}
\begin{IEEEproof}
	See Appendix.
\end{IEEEproof}

The NNI property of a SISO HIGS is shown in the following theorem.
\begin{theorem}\label{theorem:HIGS_NNI}
    Consider a SISO HIGS as in (\ref{eq:HIGS}), then the HIGS is an NNI system from input $e$ to the output $u$ with the positive definite storage function formulated as
	\begin{equation*}
		V(x_h) = \frac{1}{2k_h}x_h^2
	\end{equation*}
	satisfying
	\begin{equation}\label{eq:theorem_Vdot_HIGS}
	    \dot{V}(x_h) \leq e\dot{u}.
	\end{equation}
\end{theorem}

\begin{IEEEproof}
	The storage function $V(x_h)$ is positive definite since $k_h>0$. Here, we prove that (\ref{eq:theorem_Vdot_HIGS}) holds in both integrator and gain modes. Taking the time derivative of $V(x_h)$, we have that
	\begin{equation*}
	    \dot{V}(x_h) = \frac{1}{k_h}x_h\dot{x}_h.
	\end{equation*}
	\textit{\textbf{Case 1. }} The HIGS operates in the integrator mode; i.e., $(e,u,\dot{e}) \in \mathcal{F}_1$. In this case, we have that $\dot{x}_h = \omega_he$. Therefore, $\dot{V}$ is obtained as
    \begin{align}\label{eq:Vdot_integrator_mode}
	    \dot{V}(x_h) &= \frac{1}{k_h}\omega_h ex_h \nonumber\\
	    &\leq \omega_he^2 = e\dot{u}.
	\end{align}
	where the inequality follows from Lemma \ref{lemma:F implication}.\\
	\textit{\textbf{Case 2. }} The HIGS operates in the gain mode; i.e., $(e,u,\dot{e}) \in \mathcal{F}_2$. In this case, we have that $u = x_h = k_he$. Therefore,
	\begin{equation}\label{eq:Vdot_gain_mode}
	    \dot{V}(x_h) = \frac{1}{k_h}k_he\dot{x}_h = e\dot{u}.
	\end{equation}
According to (\ref{eq:Vdot_integrator_mode}) and (\ref{eq:Vdot_gain_mode}), and Definition~\ref{def:nonlinear_NI}, the HIGS is an NNI system.
\end{IEEEproof}

\subsection{Stability for the interconnection of a SISO NI system and a HIGS}
\begin{figure}[h!]
\centering
\psfrag{in_0}{$r=0$}
\psfrag{in_1}{$u$}
\psfrag{y_1}{$y$}
\psfrag{e}{\hspace{0.08cm}$e$}
\psfrag{x_h}{$x_h$}
\psfrag{plant}{$G(s)$}
\psfrag{HIGS}{\hspace{-0.25cm} HIGS $\mc H$}
\psfrag{+}{\small$+$}
\includegraphics[width=8cm]{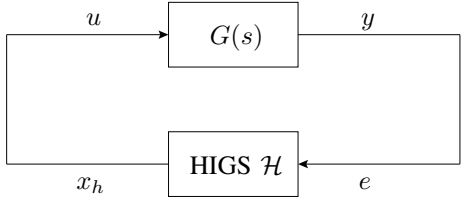}
\caption{Closed-loop interconnection of a linear NI system and a HIGS.}
\label{fig:interconnection}
\end{figure}

Consider the interconnection of a SISO linear NI plant $G(s)$ and a HIGS controller $\mc H$ as shown in Fig.~\ref{fig:interconnection}. We analyze the stability of the closed-loop system in the following. Note that here and also in Sections \ref{sec:MIMO_HIGS}, \ref{sec:cascaded HIGS} and \ref{sec:example}, HIGS controllers are applied in positive feedback, according to the control framework used in NI systems theory \cite{lanzon2008stability,shi2023output}.

Consider a SISO NI system with the transfer function matrix $G(s)$ and the minimal realization:
\begin{subequations}\label{eq:G(s)}
\begin{align}
\dot x =&\ Ax+Bu,\label{eq:G(s) state equation}\\
y =&\ Cx,
\end{align}
\end{subequations}
where $x\in \mathbb R^n$, $u,y \in \mathbb R$ are the state, input and output of the system, respectively. Here, $A\in \mathbb R^{n\times n}$, $B\in \mathbb R^{n\times 1}$ and $C\in \mathbb R^{1\times n}$.

\begin{theorem}\label{theorem:stability of single interconnection}
Consider the SISO minimal linear NI system (\ref{eq:G(s)}). There exists a HIGS element $\mc H$ of the form (\ref{eq:HIGS}) such that the closed-loop interconnection of the system (\ref{eq:G(s)}) and the HIGS $\mc H$ as shown in Fig.~\ref{fig:interconnection} is asymptotically stable.
\end{theorem}
\begin{IEEEproof}
This result is a special case of Theorem \ref{thm:Multi-HIGS stability}. See the proof of Theorem \ref{thm:Multi-HIGS stability}. Also, the proof for this special case is presented in \cite{shi2022negative}.

\end{IEEEproof}


\section{MULTI-HIGS}\label{sec:MIMO_HIGS}
In this section, we provide a description of multi-HIGS and show that a multi-HIGS is an NNI system. Also, we prove that for any MIMO linear NI system, there exists a multi-HIGS controller such that their closed-loop interconnection is asymptotically stable.
\subsection{Description of Multi-HIGS}
\begin{figure}[h!]
\centering
\psfrag{e_1}{$e_1$}
\psfrag{e_2}{$e_2$}
\psfrag{e_n}{$e_N$}
\psfrag{udd}{$\vdots$}
\psfrag{u_1}{$u_1$}
\psfrag{u_2}{$u_2$}
\psfrag{u_n}{$u_N$}
\psfrag{odd}{$\vdots$}
\psfrag{ddd}{\hspace{0.1cm}$\vdots$}
\psfrag{H_p1}{$\mc H_1$}
\psfrag{H_p2}{$\mc H_2$}
\psfrag{H_pn}{$\mc H_N$}
\psfrag{H_0}{$\widehat {\mc H}$}
\includegraphics[width=9.5cm]{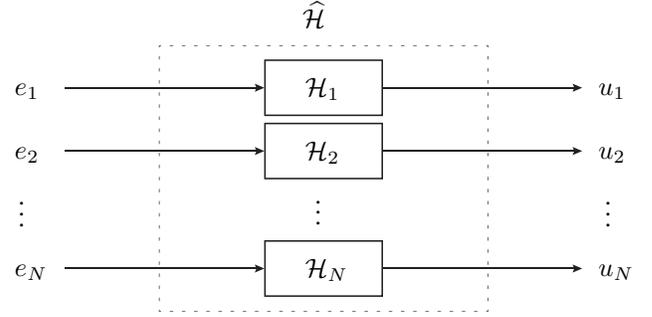}
\caption{A multi-HIGS $\widehat {\mc H}$, which is constructed by connecting N HIGS of the form (\ref{eq:HIGS}) in parallel.}
\label{fig:HIGS_MIMO}
\end{figure}

Consider $N$ HIGS of the form (\ref{eq:HIGS}) with different integrator frequencies $\omega_{h,1},\cdots, \omega_{h,N}$ and gain values $k_{h,1},\cdots,k_{h,N}$ connected in parallel as shown in Fig.~\ref{fig:HIGS_MIMO}. The HIGS are denoted by $\mc H_1, \mc H_2, \cdots,\mc H_N$ while their inputs, outputs and states are denoted by $e_1,e_2,\cdots,e_N$, $u_1,u_2,\cdots,u_N$ and $x_{h,1},x_{h,2},\cdots,x_{h,N}$, respectively. The entire system denoted by $\widehat {\mc H}$ is called a multi-HIGS (see also \cite{Achten_HIGS_Skyhook_thesis_2020}). The input, output and state of the system $\widehat {\mc H}$ are
\begin{equation}\label{eq:MIMO_HIGS_input}
	E_h = \left[\begin{matrix} 
 	e_1& e_2 &\cdots & e_N
 \end{matrix}\right]^T,
\end{equation}
\begin{equation}\label{eq:MIMO_HIGS_output}
	U_h = \left[\begin{matrix} 
 	u_1&u_2&\cdots & u_N
 \end{matrix}\right]^T,
\end{equation}
and
\begin{equation}\label{eq:MIMO_HIGS_state}
X_h = \left[\begin{matrix} 
 	x_{h,1}& x_{h,2}& \cdots & x_{h,N}
 \end{matrix}\right]^T,
\end{equation}
respectively.

The system can also be described by the following equations:
	\begin{equation}\label{eq:multi-HIGS}
		\widehat {\mathcal{H}}:
		\begin{cases}
			e_i = \theta_i^TE_h,\\
			\dot{x}_{h,i} = \omega_{h,i}e_i , & \text{if}\, (e_i,x_{h,i},\dot{e}_i) \in \mathcal{F}_{1,i}\\
			x_{h,i} = k_{h,i}e_i, & \text{if}\, (e_i,x_{h,i},\dot{e}_i) \in \mathcal{F}_{2,i}\\
			X_h = \left[\begin{matrix}
				x_{h,1},x_{h,2},\cdots, x_{h,N}
			\end{matrix}\right]^T,\\
			U_h = X_h,
		\end{cases}
	\end{equation}
where $\theta_i\in \mathbb R^N$ is a standard unit vector. And $\mc F_{1,i}$ and $\mc F_{2,i}$ are given by:
\begin{align*}
	\mathcal{F}_{1,i}& := \mathcal{F}_i \setminus \mathcal{F}_{2,i};\\
	\mathcal{F}_{2,i}& := \{(e_i,x_{h,i},\dot{e}_i) \in \mathbb{R}^3 | x_{h,i} = k_{h,i}e_i\\
& \qquad \qquad \qquad \qquad \qquad \qquad \quad \text{and}\  \omega_{h,i} e_i^2 > k_{h,i} e_i\dot{e_i}\},
\end{align*}
where
\begin{equation}\label{eq:Fi}
	\mathcal{F}_i := \{ (e_i,x_{h,i},\dot{e}_i) \in \mathbb{R}^3 |\, e_ix_{h,i} \geq \frac{1}{k_{h,i}}x_{h,i}^2\}.
\end{equation}

\subsection{NNI Property of multi-HIGS}
Consider the system $\widehat{\mc H}$ in Fig.~\ref{fig:HIGS_MIMO}. It is shown in the following that if for all $i=1,2,\cdots,N$, the system $\mc H_i$ is NNI, then the system $\widehat{\mc H}$ is also NNI.
\begin{lemma}\label{lemma:parallel NI}
Consider $N$ NNI systems connected in parallel with inputs $u_1,u_2,\cdots,u_N$, outputs $y_1,y_2,\cdots,y_N$ and states $x_1,x_2,\cdots,x_N$. They have storage functions $V_1(x_1),V_2(x_2),\cdots,V_N(x_N)$ that satisfy Definition \ref{def:nonlinear_NI}; i.e., $\dot V_i(x_i)\leq u_i(t)^T\dot y_i(t)$, $\forall t\geq 0$, $\forall i = 1,2,\cdots,N$. Then the nonlinear system having input $U = \left[\begin{matrix} 
 	u_1^T&u_2^T&\cdots & u_N^T
 \end{matrix}\right]^T$, output $U = \left[\begin{matrix} 
 	y_1^T&y_2^T&\cdots & y_N^T
 \end{matrix}\right]^T$ and state $X = \left[\begin{matrix} 
 	x_1^T&x_2^T&\cdots & x_N^T
 \end{matrix}\right]^T$ is also NNI with the storage function
 \begin{equation}\label{eq:sf_parallel_NI}
 \widehat V(X) = \Sigma_{i=1}^N V_i(x_i).
 \end{equation}
\end{lemma}
\begin{IEEEproof}
See Appendix.
\end{IEEEproof}

\begin{theorem}\label{theorem:multi-HIGS NNI}
Consider the multi-HIGS $\widehat {\mc H}$ represented by (\ref{eq:multi-HIGS}), which is also shown in Fig.~\ref{fig:HIGS_MIMO}. The system $\widehat {\mc H}$ with input $E_h$, output $U_h$ and state $X_h$, defined in (\ref{eq:MIMO_HIGS_input}), (\ref{eq:MIMO_HIGS_output}) and (\ref{eq:MIMO_HIGS_state}) respectively, is an NNI system with the storage function
\begin{equation*}
\widehat V_h(X_h) = \Sigma_{i=1}^N V_i(x_{h,i})=\frac{1}{2}X_h^TK_h^{-1}X_h,
\end{equation*}
where $K_h = diag\{k_{h,1},k_{h,2},\cdots,k_{h,N}\}$. Here, $e_i$, $u_i$, $x_{h,i}$ and $V_i(x_{h,i})$ are the input, output, state and storage function of the $i$-th SISO HIGS $\mc H_i$, respectively, $i=1,2,\cdots,N$.
\end{theorem}
\begin{IEEEproof}
The proof follows directly from Lemma \ref{lemma:parallel NI} and the NNI property of SISO HIGS of the form (\ref{eq:HIGS}), as given in Theorem \ref{theorem:HIGS_NNI}. Specifically, we have
\begin{equation}\label{eq:multi-HIGS NNI ineq}
	\dot {\widehat V}_h(X_h) = \Sigma_{i=1}^N \dot V_i(x_{h,i}) \leq \Sigma_{i=1}^N e_i \dot u_i = E_h^T \dot U_h,
\end{equation}
which satisfies Definition \ref{def:nonlinear_NI}.	
\end{IEEEproof}

\subsection{Stability of the interconnection of a MIMO linear NI system and a multi-HIGS}
\begin{figure}[h!]
\centering
\psfrag{in_0}{$r=0$}
\psfrag{in_1}{$u$}
\psfrag{y_1}{$y$}
\psfrag{e}{$E_h$}
\psfrag{x_h}{$X_h$}
\psfrag{plant}{$G(s)$}
\psfrag{HIGS}{\hspace{-0.1cm}HIGS $\widehat{\mc H}$}
\psfrag{+}{\small$+$}
\includegraphics[width=8.5cm]{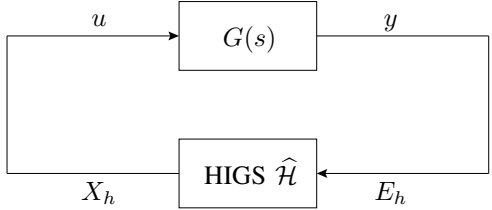}
\caption{Closed-loop interconnection of a MIMO linear NI plant $G(s)$ and a multi-HIGS controller $\widehat {\mc H}$.}
\label{fig:MIMO interconnection}
\end{figure}

\begin{lemma}\label{lemma:MIMO lossless condition}
Consider the multi-HIGS of the form (\ref{eq:multi-HIGS}) shown in Fig.~\ref{fig:HIGS_MIMO}. If $\dot {\widehat V}_h(X_h)=E_h^T \dot U_h$, then $X_h = K_hE_h$.
\end{lemma}
\begin{IEEEproof}
See Appendix.
\end{IEEEproof}

Consider a MIMO linear NI system with the transfer function matrix $G(s)$ and the minimal realization
\begin{subequations}\label{eq:MIMO NI system}
\begin{align}
\dot x =& Ax + Bu,\\
y =& Cx,	
\end{align}
where $u,y\in \mathbb R^N$ and $x\in \mathbb R^n$ are the input, output and state of the system, respectively. Here $A\in \mathbb R^{n\times n}$, $B\in \mathbb R^{n\times N}$ and $C\in \mathbb R^{N\times n}$.
\end{subequations}

\begin{theorem}\label{thm:Multi-HIGS stability}
Consider the MIMO minimal linear NI system (\ref{eq:MIMO NI system}). There always exists a multi-HIGS $\widehat {\mc H}$, given in (\ref{eq:multi-HIGS}) and Fig.~\ref{fig:HIGS_MIMO}, such that the closed-loop interconnection of the system (\ref{eq:MIMO NI system}) and the multi-HIGS $\widehat {\mc H}$ as shown in Fig.~\ref{fig:MIMO interconnection} is asymptotically stable.
\end{theorem}
\begin{IEEEproof}
Since the system (\ref{eq:MIMO NI system}) is minimal and NI, then according to Lemma \ref{lemma:NI}, we have that $\det A \neq 0$ and there exists $Y=Y^T>0$, $Y\in \mathbb R^{n\times n}$ such that
\begin{equation}\label{eq:NI lemma relations MIMO}
	AY+YA^T\leq 0,\qquad \textnormal{and} \qquad B+AYC^T=0.
\end{equation}	
Using Lyapunov's direct method, let the storage function of the closed-loop interconnection be
\begin{align*}
W(x,X_h)=&\frac{1}{2}x^TY^{-1}x+\frac{1}{2}X_h^TK_h^{-1}X_h-X_h^TCx\notag\\
=& \frac{1}{2}\left[\begin{matrix}x^T & X_h^T\end{matrix}\right]\left[\begin{matrix}Y^{-1}&-C^T\\-C&K_h^{-1}\end{matrix}\right]\left[\begin{matrix}x \\ X_h\end{matrix}\right].
\end{align*}
Using Schur complement theorem, $W(x,X_h)>0$ for all $(x,X_h)\neq (0,0)$ if
\begin{equation}\label{eq:W pd initial MIMO}
	K_h^{-1}-CYC^T>0.
\end{equation}
Using (\ref{eq:NI lemma relations MIMO}), we have that $CYC^T=-CA^{-1}B=G(0)$, where $G(s)=C(sI-A)^{-1}B$ is the transfer function matrix of the system (\ref{eq:MIMO NI system}) Then, (\ref{eq:W pd initial MIMO}) can be written as
\begin{equation*}
	K_h^{-1}-G(0)>0.
\end{equation*}
Since both $Y^{-1}$ and $K_h^{-1}$ are positive definite, the condition (\ref{eq:W pd initial MIMO}) is equivalent to
\begin{equation}\label{eq:W pd condition v2}
	Y^{-1}-C^TK_hC>0.
\end{equation}
Take the time derivative of $W(x,X_h)$, we have
\begin{align*}
\dot W&(x,X_h)\notag\\
=& x^TY^{-1}\dot x+X_hK_h^{-1}\dot X_h-\dot X_h^TC x-X_h^TC\dot x\notag\\
=& \left(x^TY^{-1}-X_hC\right)\dot x+\dot X_h^T\left(K_h^{-1}X_h-C x\right)\notag\\
=& \left(x^TY^{-1}-uC\right)\dot x+\dot X_h^T\left(K_h^{-1}X_h-E_h\right)\notag\\
=&\left(x^TY^{-1}+uB^TA^{-T}Y^{-1}\right)\dot x+\dot X_h^T\left(K_h^{-1}X_h-E_h\right)\notag\\
=&\left(x^TA^T+uB^T\right)(A^{-T}Y^{-1})\dot x+\dot X_h^T\left(K_h^{-1}X_h-E_h\right)\notag\\
=&\frac{1}{2}\dot x^T (A^{-T}Y^{-1}+Y^{-1}A^{-1})\dot x+\dot X_h^T\left(K_h^{-1}X_h-E_h\right),
\end{align*}
where $u = X_h$ and $E_h=y=Cx$ are also used. We have that $\dot X_h^T\left(K_h^{-1}X_h-E_h\right) = \dot {\widehat V}_h(X_h)-E_h^T\dot U_h\leq 0$ and equality holds only if $X_h=K_hE_h$ according to Lemma \ref{lemma:MIMO lossless condition}. We also have that $\frac{1}{2}\dot x^T (A^{-T}Y^{-1}+Y^{-1}A^{-1})\dot x\leq 0$ because $A^{-T}Y^{-1}+Y^{-1}A^{-1}\leq 0$ according to (\ref{eq:NI lemma relations MIMO}). Therefore, $\dot W(x,X_h)\leq 0$ and the equality $\dot W(x,X_h)=0$ holds only if $X_h=K_hE_h$ and $\dot x^T (A^{-T}Y^{-1}+Y^{-1}A^{-1})\dot x=0$. We apply LaSalle's invariance principle in the following to prove that there exist $k_{h,i}$ and $\omega_{h,i}$ $(i=1,2,\cdots,N)$ such that the closed-loop system is asymptotically stable. We only consider the case that $x\neq 0$. The function $\dot W(x,X_h)$ stays at zero only if $X_h\equiv K_hE_h$ and $\dot x^T (A^{-T}Y^{-1}+Y^{-1}A^{-1})\dot x\equiv 0$. The condition $X_h\equiv K_hE_h$ implies that
\begin{equation}\label{eq:HIGS dot W =0 relation MIMO}
	u\equiv K_hy \equiv K_hCx,
\end{equation}
where the system settings $u=U_h=X_h$ and $E_h=y=Cx$ as shown in Fig.~\ref{fig:MIMO interconnection} are also used. In the sequel, we have that
\begin{equation}\label{eq:dot x and x relationship}
	\dot x \equiv  A x+B u \equiv  Ax+BK_hCx \equiv (A+BK_hC)x.
\end{equation}
According to (\ref{eq:NI lemma relations MIMO}), we have that
\begin{align*}
	A+BK_hC = A-AYC^TK_hC=AY(Y^{-1}-C^TK_hC),
\end{align*}
which is nonsingular due to (\ref{eq:W pd condition v2}) and the non-singularity of the matrices $A$ and $Y$. Therefore, according to (\ref{eq:dot x and x relationship}), $\dot x \neq 0$ for any nonzero $x$. Also, $\dot x$ cannot remain a constant vector because
\begin{equation*}
	\ddot x = (A+BK_hC)\dot x\neq 0,
\end{equation*}
for any nonzero $\dot x$. The condition $\dot x^T (A^{-T}Y^{-1}+Y^{-1}A^{-1})\dot x\equiv 0$ implies that $\dot x$ must stay in the null space of $A^{-T}Y^{-1}+Y^{-1}A^{-1}$. 

With the above information about $\dot x$ known in the case that $\dot W(x,X_h)\equiv 0$, we now prove that $\dot W(x,X_h)$ cannot remain zero forever. We first prove by contradiction that none of the single HIGS can stay in integrator mode unless both of the HIGS input and output remain zero. Suppose that there is a HIGS $\mc H_i$ staying in the integrator mode $\mc F_{1,i}$. Then we have that $\dot x_{h,i} = \omega_{h,i}e_i=\omega_{h,i}y_i$ according to (\ref{eq:multi-HIGS}), and $x_{h,i} = k_{h,i} y_i$ according to (\ref{eq:HIGS dot W =0 relation MIMO}). Here, $y_i$ denotes the $i$-th output element in $y$. Therefore, we have that
\begin{equation}\label{eq:y dot y relationship for contradiction}
	\dot x_{h,i}=\omega_{h,i}y_i=k_{h,i} \dot y_i
\end{equation}
over a nonzero time interval $[t_a,t_b]$ where $t_a<t_b$. With $\omega_{h,i}>0$ chosen, for nonzero $y_i(t_a)$, (\ref{eq:y dot y relationship for contradiction}) implies that $y_i(t) = y_i(t_a)exp(\frac{\omega_{h,i}}{k_{h,i}}t)$ for $t\in [t_a,t_b]$. This means that the closed-loop system is unstable, which contradicts its Lyapunov stability proved above. For HIGS with an input $e_j=y_j$ that does not remain zero, it must stay in the gain mode $\mc F_{2,j}$. Now we prove that we can force it to exit the gain mode by choosing suitable HIGS parameters. Now we consider the condition $\omega_{h,j}e_j^2>k_{h,j}e_j\dot e_j$ in $\mc F_{2,j}$. This condition cannot be satisfied for all HIGS $\mc H_j$ in gain mode over time via satisfying $e_j\dot e_j<0$ because then $\dot V_{h,j}\leq e_j \dot x_{h,j}=k_{h,j}e_j\dot e_j<0$, where $V_{h,j}$ is the storage function of the HIGS $\mc H_j$. Hence $y_j=e_j=\frac{1}{k_{h,j}}x_{h,j}$ will converge to zero. This implies that eventually $y=0$, which is not the case considered here. Considering those $e_j$ that eventually satisfies $e_j\dot e_j>0$, since the trajectories of $\dot e_j$ and $e_j$ in gain mode is independent of $\omega_{h,j}$, we can always choose sufficiently small $\omega_{h,j}>0$ such that $\omega_{h,j}e_j^2<k_{h,j}e_j\dot e_j$, in order to violate the condition $\mc F_{2,j}$ for some $\mc H_j$ in gain mode. These $\mc H_j$ will then enter integrator mode for at least some finite time. Thus, we can eventually force all HIGS to switch into integrator mode and have their inputs remaining zero, except for those staying in the gain mode by satisfying $\mc F_{2,j}$ via $\dot e_j\equiv 0$. In this case, $\dot y \equiv 0$, which implies that $\dot x \equiv 0$ according to observability. This contradicts with the fact that $\dot x$ cannot remain zero, which has been proved above. We conclude that $\dot W(x,X_h)\equiv 0$ will eventually be violated, and $W(x,X_h)$ will decrease monotonically until it reaches zero.

\end{IEEEproof}

\begin{remark}
	The multi-HIGS model (\ref{eq:multi-HIGS}) allows the integrator frequencies $\omega_{h,i}$ to be zero. However, we need to choose strictly positive integrator frequencies in some cases. For example, a lossless NI plant $G(s)$ cannot be stabilized by a pure gain feedback. Hence, we need $\omega_{h,i}>0$ at least for some $i$, or even for all $i$, to ensure the HIGS controllers will work properly. Similar remarks also apply to the results in Theorems \ref{theorem:stability of single interconnection} and \ref{theorem:cascade stability}.
\end{remark}

\section{THE CASCADE OF TWO HIGS}\label{sec:cascaded HIGS}
In this section, the cascade of two HIGS elements is considered as a controller for SISO linear NI systems. we prove that the closed-loop system featuring the cascaded HIGS is asymptotically stable.
\subsection{System Description}
\begin{figure}[h!]
\centering
\psfrag{U_p}{$e_1$}
\psfrag{Y_p}{$x_1$}
\psfrag{U_nc}{\hspace{0.3cm}$e_2$}
\psfrag{Y_nc}{$x_2$}
\psfrag{H_p}{\hspace{-0.4cm}\small HIGS $\mc H_1$}
\psfrag{H_nc}{\hspace{-0.4cm}\small HIGS $\mc H_2$}
\includegraphics[width=8.5cm]{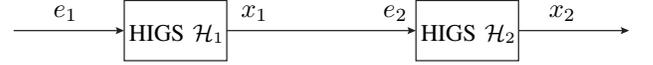}
\caption{A cascade of the HIGS $\mc H_1$ and the HIGS $\mc H_2$. The output $x_1$ of $\mc H_1$ is fed into $\mc H_2$ as its input $e_2$.}
\label{fig:cascade}
\end{figure}

We first provide a description for the cascaded HIGS. As shown in Fig.~\ref{fig:cascade}, it is a simple open-loop interconnection of two HIGS where the output of one HIGS is used as the output of the other HIGS.

The system model of the two HIGS system $\mc H_1$ and $\mc H_2$ are as follows:
\begin{align}\label{eq:HIGS H_1}
\mc H_1: \quad 
\begin{cases}
\dot x_1 = \omega_1e_1, & \text{if} (e_1,y_1,\dot e_1)\in\mathcal F_1\\
x_1 = k_1e_1, & \text{if} (e_1,y_1,\dot e_1)\in \mathcal F_2\\
y_1 = x_1,	
\end{cases}
\end{align}
where $x_1,e_1,y_1\in \mathbb R$ are the state, input and output of the system, respectively. Here, $\omega_1\in [0,\infty)$ and $k_1\in (0,\infty)$ are system parameters. And we have that
\begin{equation}\label{eq:F for H1}
\mathcal F= \left\{(e_1,y_1,\dot e_1)\in\mathbb R^3|e_1y_1\geq \frac{1}{k_1}y_1^2\right\},	
\end{equation}
\begin{equation*}
\mc F_1=\mc F\backslash \mc F_2,
\end{equation*}
\begin{equation}\label{eq:F2 for H1}
\mc F_2=\left\{(e_1,y_1,\dot e_1)\in \mc F|y_1=k_1e_1\ \text{and}\ \omega_1e_1^2>k_1\dot e_1e_1\right\}.
\end{equation}

\begin{align}\label{eq:HIGS H_2}
\mc H_2: \quad 
\begin{cases}
\dot x_2 = \omega_2e_2, & \text{if} (e_2,y_2,\dot e_2)\in\widetilde{\mathcal  F}_1\\
x_2 = k_2e_2, & \text{if} (e_2,y_2,\dot e_2)\in \widetilde{\mathcal  F}_2\\
y_2 = x_2,	
\end{cases}
\end{align}
where $x_2,e_2,y_2\in \mathbb R$ are the state, input and output of the system, respectively. Here, $\omega_2\in [0,\infty)$ and $k_2\in (0,\infty)$ are system parameters. And we have that
\begin{equation}\label{eq:F for H2}
\widetilde{\mathcal  F}= \left\{(e_2,y_2,\dot e_2)\in\mathbb R^3|e_2y_2\geq \frac{1}{k_2}y_2^2\right\},\end{equation}
\begin{equation*}
\widetilde{\mc F}_1=\widetilde{\mc F}\backslash \widetilde{\mc F}_2,
\end{equation*}
\begin{equation}\label{eq:F2 for H2}
\widetilde{\mc F}_2=\left\{(e_2,y_2,\dot e_2)\in \widetilde{\mc F}|y_2=k_2e_2 \ \text{and} \ \omega_2e_2^2>k_2\dot e_2e_2\right\}.
\end{equation}
The interconnection can be described by the equation
\begin{equation*}\label{eq:cascaded HIGS setting}
	e_2 = y_1.
\end{equation*}

\subsection{NNI Property of the Cascade of Two HIGS}

The following lemma is required in the presentation of the main results.
\begin{lemma}\label{lemma:equality holds condition in F}
Consider the HIGS of the form (\ref{eq:HIGS}). Suppose $ex_h=\frac{1}{k_h}x_h^2$ over a time interval $[t_a,t_b]$, where $t_a<t_b$, then $x_h = k_he$ for all $t\in [t_a,t_b]$.
\end{lemma}
\begin{IEEEproof}
See Appendix.
\end{IEEEproof}
Note that Lemma \ref{lemma:equality holds condition in F} is nontrivial since the equation $ex_h=\frac{1}{k_h}x_h^2$ has two solutions $x_h=0$ and $x_h = k_he$.

\begin{theorem}\label{theorem:cascade HIGS NI}
Consider two HIGS $\mc H_1$ and $\mc H_2$ having the system models (\ref{eq:HIGS H_1}) and (\ref{eq:HIGS H_2}), respectively. Suppose $k_2 \omega_1 \leq k_1 \omega_2$, then the cascade of $\mc H_1$ and $\mc H_2$ as shown in Fig.~\ref{fig:cascade} is an NNI system with the Lyapunov storage function
\begin{equation}\label{eq:V for cascade}
V(x_1,x_2) = ax_1^2+\frac{k_2-2ak_1}{2k_1k_2^2}x_2^2,	
\end{equation}
satisfying
\begin{equation}\label{eq:NI ineq cascade}
\dot V(x_1,x_2) \leq e_1 \dot x_2.	
\end{equation}
Here, $0<a<\frac{k_2}{2k_1}$ is a parameter. Moreover, if $\dot V(x_1,x_2)=e_1\dot x_2$ over a time interval $[t_a,t_b]$, where $t_a<t_b$, then for all $t\in [t_a,t_b]$ we have that $x_1 = k_1e_1$ and $x_2 = k_2e_2=k_2k_1e_1$. 
\end{theorem}
\begin{IEEEproof}
Since $0<a<\frac{k_2}{2k_1}$, we have that $V(x_1,x_2)$ in (\ref{eq:V for cascade}) is positive definite. Because both $\mc H_1$ and $\mc H_2$ have two modes, i.e., integrator mode and gain mode, the cascaded system has four modes. We prove in the following that (\ref{eq:NI ineq cascade}) holds in these four modes. Note that $e_2 = x_1$ will be used in the following.

\textit{\textbf{Case 1.}} $\mc H_1$ in $\mc F_1$ and $\mc H_2$ in $\widetilde {\mc F}_1$. According to (\ref{eq:F for H1}) and (\ref{eq:F for H2}), we have that
\begin{align}
e_1x_1\geq & \frac{1}{k_1}x_1^2,\label{eq:case1 condition1}\\
x_1x_2 \geq & \frac{1}{k_2}x_2^2 \implies x_1x_2\leq k_2x_1^2,	\label{eq:case1 condition2}  
\end{align}
where the deduction in (\ref{eq:case1 condition2}) uses Lemma \ref{lemma:F implication}. Since $k_2 \omega_1 \leq k_1 \omega_2$ and $0<a<\frac{k_2}{2k_1}$, we have that $2a\omega_1-\omega_2<0$ and $k_2-2ak_1>0$. Taking the time derivative of $V(x_1,x_2)$, we get
\begin{align}
\dot V&(x_1,x_2)-e_1\dot x_2\notag\\
=& 2ax_1\dot x_1+ \frac{k_2-2ak_1}{k_1k_2^2}x_2\dot x_2-e_1\dot x_2\notag\\
=& 	2ax_1\omega_1e_1+ \frac{k_2-2ak_1}{k_1k_2^2}x_2\omega_2x_1-\omega_2e_1x_1\notag\\
=& 	(2a\omega_1-\omega_2)e_1x_1+ \frac{k_2-2ak_1}{k_1k_2^2}\omega_2x_1x_2 \notag\\
\leq & (2a\omega_1-\omega_2)\frac{1}{k_1}x_1^2+\frac{k_2-2ak_1}{k_1k_2^2}\omega_2k_2x_1^2\notag\\
= & \frac{1}{k_1}x_1^2\left(2a\omega_1-\omega_2+\frac{k_2-2ak_1}{k_2}\omega_2\right)\notag\\
= & \frac{2a}{k_1}x_1^2\left(\omega_1-\frac{k_1}{k_2}\omega_2\right)\notag\\
\leq & 0,\label{eq:Case1}
\end{align}
where the first inequality uses (\ref{eq:case1 condition1}), (\ref{eq:case1 condition2}) and also the fact that $2a\omega_1-\omega_2<0$. This implies in (\ref{eq:Case1}), $\dot V(x_1,x_2)-e_1\dot x_2=0$ is possible only if equalities in (\ref{eq:case1 condition1}) and (\ref{eq:case1 condition2}) hold. If the equalities hold over the time interval $[t_a,t_b]$, then according to Lemmas \ref{lemma:F implication} and \ref{lemma:equality holds condition in F}, we have that $x_1 = k_1e_1$ and $x_2 = k_2e_2$ in $[t_a,t_b]$. That is $x_2 = k_1k_2e_1$.

\textit{\textbf{Case 2.}} $\mc H_1$ in $\mc F_2$ and $\mc H_2$ in $\widetilde{\mc F}_2$.
Take the time derivative of $V(x_1,x_2)$:
\begin{align*}
\dot V&(x_1,x_2)-e_1\dot x_2\notag\\
=& 2ax_1\dot x_1+ \frac{k_2-2ak_1}{k_1k_2^2}x_2\dot x_2-e_1\dot x_2\notag\\
=& 2ak_1^2e_1\dot e_1 + \frac{k_2-2ak_1}{k_1k_2^2}k_1^2k_2^2e_1\dot e_1-k_1k_2e_1\dot e_1\notag\\
=& e_1\dot e_1\left(2ak_1^2+\frac{k_2-2ak_1}{k_1k_2^2}k_1^2k_2^2-k_1k_2\right)\notag\\
=& 0.
\end{align*}
In Case 2, $x_1= k_1e_1$ and $x_2 = k_2e_2$ automatically holds.

\textit{\textbf{Case 3.}} $\mc H_1$ in $\mc F_1$ and $\mc H_2$ in $\widetilde {\mc F}_2$. According to (\ref{eq:F for H1}), we have that
\begin{equation}
e_1x_1\geq \frac{1}{k_1}x_1^2\implies e_1x_1\leq k_1e_1^2,\label{eq:Case 3 implication 1}
\end{equation}
where the deduction uses Lemma \ref{lemma:F implication}. And according to (\ref{eq:F2 for H2}), we have that
\begin{equation*}
	\omega_2e_2^2> k_2\dot e_2e_2.\label{eq:Case 3 implication 2}
\end{equation*}
We take the time derivative of $V(x_1,x_2)$:
\begin{align}
\dot V&(x_1,x_2)-e_1\dot x_2\notag\\
=& 2ax_1\dot x_1+ \frac{k_2-2ak_1}{k_1k_2^2}x_2\dot x_2-e_1\dot x_2\notag\\
=& 2ax_1\omega_1e_1 + \frac{k_2-2ak_1}{k_1k_2^2}k_2x_1k_2\omega_1e_1 - e_1k_2\omega_1e_1\notag\\
=& \frac{\omega_1k_2}{k_1}x_1e_1-\omega_1k_2e_1^2\notag\\
\leq & \frac{\omega_1k_2}{k_1}k_1e_1^2-\omega_1k_2e_1^2\notag\\
=&0,\label{eq:Case3}
\end{align}
where the inequality also uses (\ref{eq:Case 3 implication 1}).
Therefore, in (\ref{eq:Case3}), $\dot V(x_1,x_2)-e_1\dot x_2=0$ over a time interval $[t_a,t_b]$ only if $x_1 = k_1e_1$ in $[t_a,t_b]$, according to Lemma \ref{lemma:F implication}. Also, as $\mc H_2$ is in $\widetilde{\mc F}_2$ mode, $x_2 = k_2e_2$ automatically holds. This implies that $x_2 = k_1k_2 e_1$.

\textit{\textbf{Case 4.}} $\mc H_1$ in $\mc F_2$ and $\mc H_2$ in $\widetilde {\mc F}_1$. According to (\ref{eq:F2 for H1}) and (\ref{eq:F for H2}) we have that
\begin{align*}
\omega_1e_1^2 > & k_1\dot e_1 e_1,\\
x_1x_2 \geq & \frac{1}{k_2}x_2^2 \implies x_1x_2\leq k_2x_1^2	 \implies e_1x_2\leq k_1k_2e_1^2.
\end{align*}
Take the time derivative of $V(x_1,x_2)$:
\begin{align*}
\dot V&(x_1,x_2)-e_1\dot x_2\notag\\
=& 2ax_1\dot x_1+ \frac{k_2-2ak_1}{k_1k_2^2}x_2\dot x_2-e_1\dot x_2\notag\\
=& 2ak_1^2e_1\dot e_1+ \frac{k_2-2ak_1}{k_1k_2^2}x_2\omega_2k_1e_1-e_1\omega_2k_1e_1\notag\\
< & 2ak_1\omega_1e_1^2 + \frac{k_2-2ak_1}{k_1k_2^2}\omega_2k_1k_1k_2e_1^2-\omega_2k_1e_1^2\notag\\
=& 2ak_1e_1^2(\omega_1-\omega_2\frac{k_1}{k_2})\notag\\
\leq & 0.
\end{align*}
In Case 4, $\dot V(x_1,x_2)<e_1\dot x_2$ and hence the case $\dot V(x_1,x_2)=e_1\dot x_2$ does not exist. We conclude that in all four cases, we have that $\dot V(x_1,x_2)\leq e_1\dot x_2$. And if $\dot V(x_1,x_2)=e_1\dot x_2$ over a time interval $[t_a,t_b]$, where $t_a<t_b$, then $x_1=k_1e_1$ and $x_2 = k_2e_2 =  k_1k_2e_1$ for all $t\in [t_a,t_b]$.
\end{IEEEproof}

\subsection{Stability for the interconnection of an NI system and two cascaded HIGS}
\begin{figure}[h!]
\centering
\psfrag{in_0}{$r=0$}
\psfrag{in_1}{$u$}
\psfrag{y_1}{$y$}
\psfrag{e}{$e$}
\psfrag{x_h}{$x_h$}
\psfrag{plant}{$G(s)$}
\psfrag{HIGS}{$HIGS$}
\psfrag{+}{\small$+$}
\psfrag{U_p}{$e_1$}
\psfrag{Y_p}{\hspace{0.1cm}$x_1$}
\psfrag{U_nc}{\hspace{0.3cm}$e_2$}
\psfrag{Y_nc}{\hspace{0.2cm}$x_2$}
\psfrag{H_p}{\hspace{0.1cm}$\mc H_1$}
\psfrag{H_nc}{\hspace{0.2cm}$\mc H_2$}
\includegraphics[width=8.5cm]{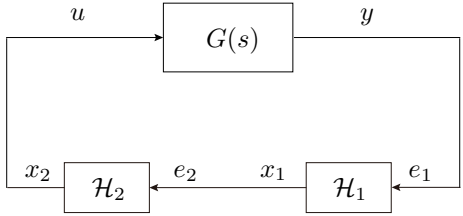}
\caption{Closed-loop interconnection of a linear NI system and the cascade of two HIGS.}
\label{fig:closed_cascade}
\end{figure}

\begin{theorem}\label{theorem:cascade stability}
	Consider the SISO minimal linear NI system (\ref{eq:G(s)}). There exist two HIGS elements $\mc H_1$ and $\mc H_2$, given in (\ref{eq:HIGS H_1}) and (\ref{eq:HIGS H_2}), respectively, such that the interconnection of the system (\ref{eq:G(s)}) and the cascade of $\mc H_1$ and $\mc H_2$ as shown in Fig.~\ref{fig:closed_cascade} is asymptotically stable.
\end{theorem}
\begin{IEEEproof}
Since the system (\ref{eq:G(s)}) is minimal and NI, then according to Lemma \ref{lemma:NI}, we have that $\det A \neq 0$ and there exists $Y=Y^T>0$, $Y\in \mathbb R^{n\times n}$ such that
\begin{equation}\label{eq:NI lemma relations2}
	AY+YA^T\leq 0,\qquad \textnormal{and} \qquad B+AYC^T=0.
\end{equation}	
Using Lyapunov's direct method, let the storage function of the closed-loop interconnection be
\begin{align}
W(x,x_1,x_2)=&\frac{1}{2}x^TY^{-1}x+V(x_1,x_2)-Cxx_2\notag\\
=&\frac{1}{2}x^TY^{-1}x+ax_1^2+\frac{k_2-2ak_1}{2k_1k_2^2}x_2^2-Cxx_2\notag\\
=& ax_1^2+ \frac{1}{2}\left[\begin{matrix}x^T & x_2\end{matrix}\right]\left[\begin{matrix}Y^{-1}&-C^T\\-C&\frac{k_2-2ak_1}{k_1k_2^2}\end{matrix}\right]\left[\begin{matrix}x \\ x_2\end{matrix}\right],\label{eq:W(x,x_1,x_2)}
\end{align}
where $V(x_1,x_2)$, as given in (\ref{eq:V for cascade}), is the storage function of the cascade of $\mc H_1$ and $\mc H_2$. Here, $a\in \mathbb R$ is a constant parameter that satisfies
\begin{equation}\label{eq:a region}
	0<a<\frac{k_2}{2k_1}.
\end{equation}
Using Schur complement theorem, $W(x,x_1,x_2)>0$ for all $(x,x_1,x_2)\neq (0,0,0)$ if
\begin{equation}\label{eq:W cascade initial}
	\frac{k_2-2ak_1}{k_1k_2^2}-CYC^T>0.
\end{equation}
Using (\ref{eq:NI lemma relations2}), we have that $CYC^T=-CA^{-1}B=G(0)$, where $G(s)=C(sI-A)^{-1}B$ is the transfer function matrix of the system (\ref{eq:G(s)}). Then, (\ref{eq:W cascade initial}) becomes
\begin{equation}\label{eq:k_1 k_2 a G(0)}
	\frac{k_2-2ak_1}{k_1k_2^2}>G(0).
\end{equation}
There exist an $a$ in the region (\ref{eq:a region}) such that (\ref{eq:k_1 k_2 a G(0)}) holds if
\begin{equation*}
	\frac{k_2}{k_1k_2^2}>G(0).
\end{equation*}
This implies that
\begin{equation*}
k_1k_2G(0)<1.	
\end{equation*}
According to Schur complement theorem, the positive definiteness of $W(x,x_1,x_2)$ in (\ref{eq:W(x,x_1,x_2)}) also implies 
\begin{equation*}
	Y^{-1}-\frac{k_1k_2^2}{k_2-2ak_1}C^TC>0,
\end{equation*}
which is equivalent to the condition (\ref{eq:W cascade initial}). Considering (\ref{eq:a region}), we have that
\begin{equation}\label{eq:PD condition in cascaded case}
	Y^{-1}-k_1k_2C^TC>Y^{-1}-\frac{k_1k_2^2}{k_2-2ak_1}C^TC>0.
\end{equation}
Take the time derivative of $W(x,x_1,x_2)$, we have
\begin{align}
\dot W&(x,x_1,x_2)\notag\\
=& x^TY^{-1}\dot x+\dot V(x_1,x_2)-C\dot xx_2-Cx\dot x_2\notag\\
=& \left(x^TY^{-1}-x_2C\right)\dot x+\dot V(x_1,x_2)-e_1\dot x_2\notag\\
=& \left(x^TY^{-1}-uC\right)\dot x+\left(\dot V(x_1,x_2)-e_1\dot x_2\right)\notag\\
=&\left(x^TY^{-1}+uB^TA^{-T}Y^{-1}\right)\dot x+\left(\dot V(x_1,x_2)-e_1\dot x_2\right)\notag\\
=&\left(x^TA^T+uB^T\right)(A^{-T}Y^{-1})\dot x+\left(\dot V(x_1,x_2)-e_1\dot x_2\right)\notag\\
=&\frac{1}{2}\dot x^T (A^{-T}Y^{-1}+Y^{-1}A^{-1})\dot x+\left(\dot V(x_1,x_2)-e_1\dot x_2\right),\label{eq:cascade dot W}
\end{align}
where $u = x_2$ and $e_1=y=Cx$ are also used. According to Theorem \ref{theorem:cascade HIGS NI}, we have that for $k_1,k_2,\omega_1,\omega_2$ satisfying $k_2 \omega_1 \leq k_1 \omega_2$, the inequality $\dot V(x_1,x_2)-e_1\dot x_2 \leq 0$ always holds. Also, $\dot x^T (A^{-T}Y^{-1}+Y^{-1}A^{-1})\dot x\leq 0$ because $A^{-T}Y^{-1}+Y^{-1}A^{-1}\leq 0$, according to (\ref{eq:NI lemma relations2}). Therefore, $\dot W(x,x_1,x_2)\leq 0$. Using LaSalle's invariance principle, $\dot W(x,x_1,x_2)$ remains zero if both $\dot x^T (A^{-T}Y^{-1}+Y^{-1}A^{-1})\dot x$ and $\dot V(x_1,x_2)-e_1\dot x_2$ remain zero. According to Theorem \ref{theorem:cascade HIGS NI}, $\dot V(x_1,x_2)-e_1\dot x_2$ remains zero only if $x_1 \equiv k_1 e_1$ and $x_2 \equiv k_2e_2 \equiv k_1k_2e_1$. That is
\begin{equation}\label{eq:u and x relation in cascaded case}
	u\equiv k_1k_2C x.
\end{equation}
We only consider the case that $x\neq 0$ in the following. This is because $x=0$ implies that $x_2=u=0$ according to (\ref{eq:u and x relation in cascaded case}). Also, $x=0$ implies $e_1=y=0$, which then implies $x_1=0$ according to (\ref{eq:F for H1}). Here, the system settings of the interconnection in Fig.~\ref{fig:closed_cascade} are also used.

In this case, the state equation (\ref{eq:G(s) state equation}) of the system $G(s)$ becomes
\begin{equation*}
	\dot x \equiv Ax+Bu \equiv Ax + Bk_1k_2Cx = (A+k_1k_2BC)x.
\end{equation*}
According to (\ref{eq:NI lemma relations2}),
\begin{equation*}
	A+k_1k_2BC=A-k_1k_2AYC^TC=AY(Y^{-1}-k_1k_2C^TC),
\end{equation*}
which is nonsingular due to (\ref{eq:PD condition in cascaded case}) and the positive definiteness of the matrices $A$ and $Y$. This implies $\dot x\neq 0$ for all $x \neq 0$. The condition $\dot x^T (A^{-T}Y^{-1}+Y^{-1}A^{-1})\dot x\equiv 0$ implies that $\dot x$ must stay in the null space of $A^{-T}Y^{-1}+Y^{-1}A^{-1}$. We now prove that $\dot W(x,x_1,x_2)$ cannot remain zero forever. First, we prove by contradiction that neither $\mc H_1$ nor $\mc H_2$ can stay in the integrator mode. Suppose $\mc H_1$ is in the integrator mode $\mc F_1$, then $\dot x_1 = \omega_1 e_1=\omega_1 y$ according to (\ref{eq:HIGS H_1}). Also, since $x_1 = k_1e_1 = k_1y$, we have that
\begin{equation}\label{eq:cascade ODE 1}
	\dot x_1 = \omega_1 y = k_1\dot y.
\end{equation}
Since the system $G(s)$ is observable and $x$ does not remain zero, then $y$ does not remain zero. Choose $\omega_1>0$, (\ref{eq:cascade ODE 1}) implies that for a nonzero time interval $[t_a,t_b]$ where $t_a<t_b$, we have
\begin{equation*}
	y(t) = y(t_a)exp(\frac{\omega_1}{k_1}t).
\end{equation*}
This contradicts the fact that the closed-loop interconnection is Lyapunov stable, as is shown by (\ref{eq:cascade dot W}). Similarly, if $\mc H_2$ is in the integrator mode, then we have 
$\dot x_2 = \omega_2 e_2 = \omega_2 k_1 y$ and $x_2 = k_1k_2e_1 = k_1k_2y$. Following a similar analysis, this also leads to a contradiction. Then we conclude that both of the HIGS $\mc H_1$ and $\mc H_2$ are in the gain mode. We prove that we can force them to eventually exit the gain mode. In this case, the condition $\omega_1 e_1^2 > k_1 e_1 \dot e_1$ is satisfied according to (\ref{eq:F2 for H2}). This condition cannot always be satisfied over time via satisfying $e_1\dot e_1<0$ because then the NI inequality $\dot V_1(x_1)\leq e_1\dot x_1 = k_1 e_1\dot e_1<0$ implies that $x_1$ converges to zero and so does $y$ since $x_1=k_1 y$. This is not the case considered here. Therefore, $e_1$ and $\dot e_1$ will eventually satisfy $e_1\dot e_1>0$ at some time. Since the trajectory of $\dot e_1$ and $e_1$ is independent of $\omega_1$, we can choose sufficiently small $\omega_1$ such that $\omega_1e_1^2<k_1e_1\dot e_1$, in order to violate the condition $\mc F_2$. Then $\mc H_1$ will enter the integrator mode for at least some finite time. Following a similar analysis, we can choose suitable $\omega_2$ to force $\mc H_2$ to eventually enter the integrator mode. As is proved above, the function $\dot W(x,x_1,x_2)$ cannot remain zero in the integrator mode. Therefore, $W(x,x_1,x_2)$ will decrease monotonically until it reaches zero.\end{IEEEproof}

\section{ILLUSTRATIVE EXAMPLE: A MEMS NANOPOSITIONER}\label{sec:example}
The NI properties of the HIGS elements shown in Theorems \ref{theorem:HIGS_NNI}, \ref{theorem:multi-HIGS NNI}, \ref{theorem:cascade HIGS NI} and the stability results shown in Theorems \ref{theorem:stability of single interconnection}, \ref{thm:Multi-HIGS stability}, \ref{theorem:cascade stability} motivate the methodology of using HIGS in positive feedback to control flexible structures with colocated force actuators and position sensors. One example of such flexible structures is the MEMS nanopositioner. In this section, we apply the methodology experimentally on a 2-DOF MEMS nanopositioner, which is a two-input two-output (TITO) linear NI system. Under the control of a TITO multi-HIGS controller in positive feedback, the MEMS nanopositioner can track a reference signal quickly and accurately.

\begin{figure}[h!]
	\centering
	\includegraphics[width=8cm]{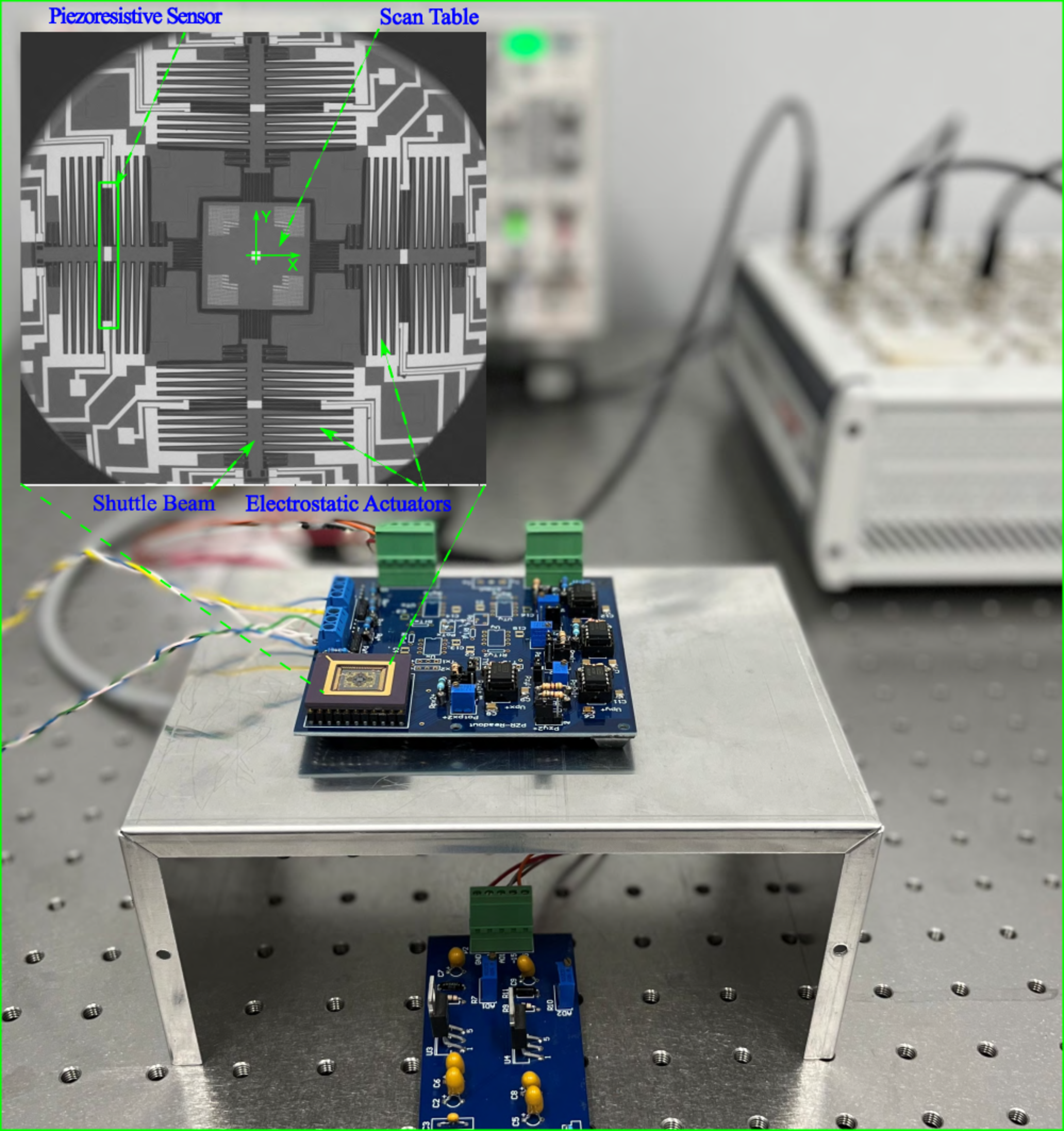}
	\caption{Experimental setup with the MEMS nanopositioner mounted on a custom-designed PCB. The close-up view shows the SEM image of the MEMS nanopositioner reported in~\cite{Maroufi_2DOF_2016}. }\label{fig:setup_SEM}
\end{figure}
\subsection{2-DOF MEMS Nanopositioner}
A 2-DOF MEMS nanopositioner is a flexible structure with colocated force actuators and position sensors, which can be regarded as an NI system \cite{Maroufi_2DOF_2016}. The nanopositioner features a stage at the center with dimensions of $1.8\, \text{mm} \times 1.8\, \text{mm}$. Four electrostatic comb-drives move the stage bidirectionally in X and Y directions.  On-chip piezoresistive sensors measure lateral displacements of the stage. The close-up view in
Fig.~\ref{fig:setup_SEM} depicts the scanning electron microscope (SEM)  image of the MEMS nanopositioner. To alleviate the quadratic nonlinearity between the induced electrostatic force and the stage displacement, this device uses a bilateral actuation mechanism. A maximum linear displacement range of  $13\, \mu\text{m}$ can be achieved in each axis. The nanopositioner was previously designed and characterized in~\cite{Maroufi_2DOF_2016} and employed as the scanner stage of a video-rate atomic force microscope~\cite{Nikooienejad_ILC_2021}.

\subsection{Frequency Response}
\begin{figure*}[h!]
	\centering
	\includegraphics[width=0.8\linewidth]{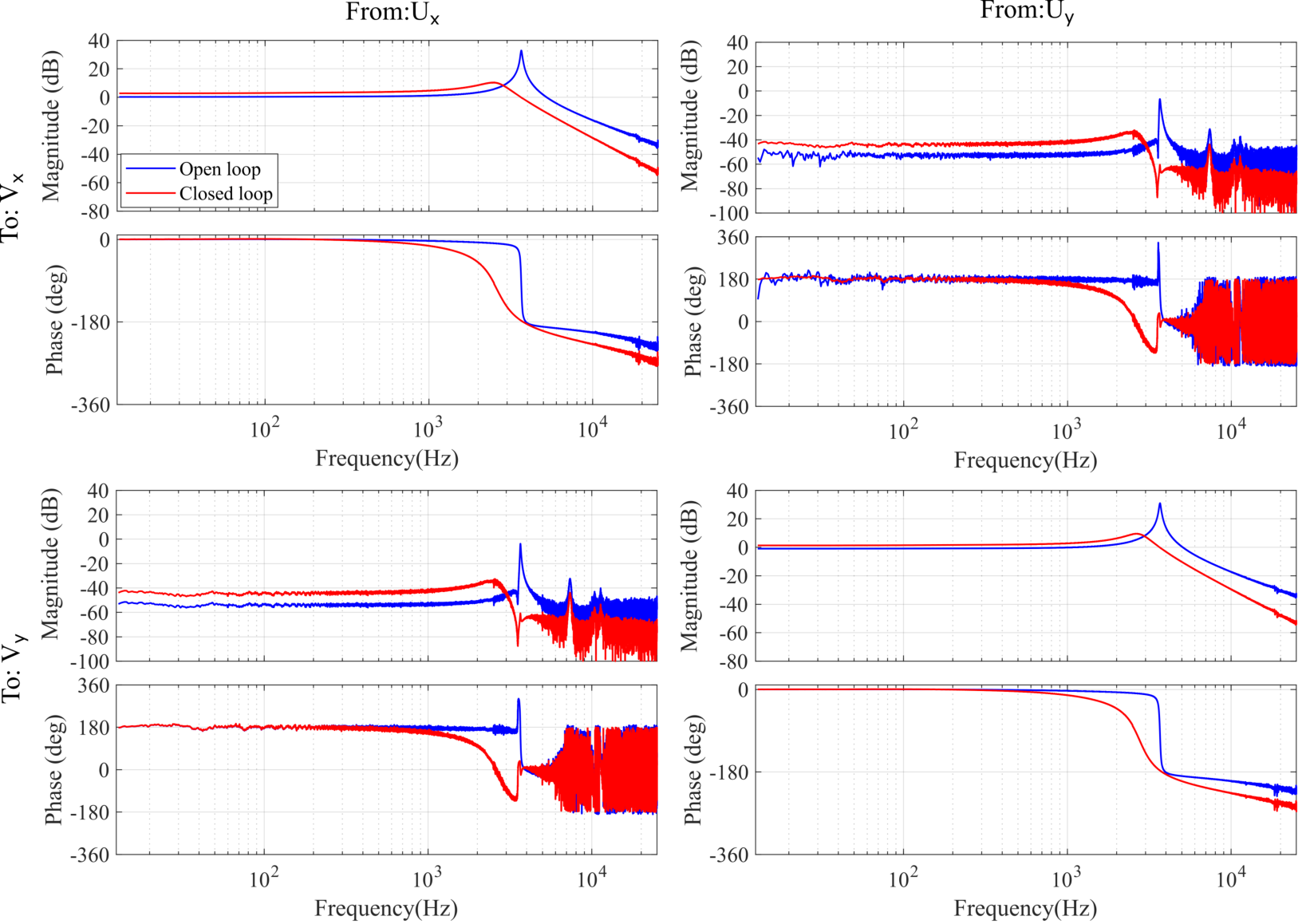}
	\caption{Frequency response of the MEMS nanopositioner in open loop and closed loop with the HIGS element in TITO configuration.}
 \label{fig:TITO_FRF}
\end{figure*}
The TITO Frequency response function of the  MEMS nanopositioner is obtained from actuator input to sensor output using an ONOSSOKI FFT Analyzer (CF-9400) with single channel excitation. For this purpose, a wideband chirp signal is applied to the actuators through a high-voltage amplifier with a gain of 20, and the frequency response of the 2-DOF nanopositioner is recorded up to $25\, \text{kHz}$. Fig.~\ref{fig:TITO_FRF} shows the measured open-loop frequency response of the device. The fundamental resonance frequencies of the X and Y axis are $3665\, \text{Hz}$ and $3680\, \text{Hz}$, respectively. We observe that the frequency responses of both axes of the MEMS nanopositioner are almost identical, with negligible cross-couplings at low frequencies. Typically, a flexure-guided nanopositioner with compatible collocated actuator-sensor pairs has NI property. Although the MEMS nanopositioner is NI in theory, it can be seen from Fig.~\ref{fig:TITO_FRF} that some high-frequency dynamics slightly violate the constraints of NI property. This is unavoidable due to the fabrication tolerances and signal conditioning in the read-out circuits, which cause the discrepancy from its ideal model. However, As discussed in~\cite{Nikooienejad_Data_Driven_2022}, the 2-DOF MEMS nanopositioner can be considered an NI system for frequencies up to 3976 Hz. We show in the following that the stability results in Theorem \ref{thm:Multi-HIGS stability} remain effective even in the presence of these spillover dynamics. The transfer function of the nanopositioner in TITO format can be described by
\begin{equation*}\label{eq:transfer_function_matrix}
G(s) = \begin{bmatrix}
G_{xx}(s) & G_{xy}(s)\\
G_{yx}(s) & G_{yy}(s)
\end{bmatrix}
\end{equation*}
where $G_{ij}(s)$ denotes the transfer function from input $j$ to the sensor output $i$. A minimal state-space realization of the transfer function matrix is obtained using the frequency response data (FRD) model from the frequency response measurement and MATLAB system identification toolbox. Accordingly, the state-space model of the nanopositioner in lateral axes can be written as
\begin{align}
    \dot{x}(t) &= Ax(t) + Bu(t),\notag \\
    y(t) &= Cx(t),\label{eq:state_space_model}
\end{align}
where
\begin{align*}
A &= \begin{bmatrix}
6989.99 & 22987.75 & 4565.46 & -1067.36\\
-25554.29 & -7675.92 & -1058.081 & -4147.69\\
1399.21	 & 505.53 &	-7758.92 &	24470.97\\
569.57 & -1001.98 &	-24272.18 &	7329.32
\end{bmatrix},\\
B &= \begin{bmatrix}
157.81 & 24.58\\
-197.95 & 12.53\\
33.58 & -89.97\\
-18.69 & -86.94
\end{bmatrix}, \\
C &= \begin{bmatrix}
-98.83 & -79.75	& -17.51 & -21.31\\
17.07 & -27.97 & -155.24 & 161.47
\end{bmatrix}.
\end{align*}
As described in~\cite{mabrok2013spectral}, the NI property of $G(s)$ can be assessed using the Hamiltonian method since $CB + B^{T}C^{T} > 0$.  According to Theorem 1 in~\cite{mabrok2013spectral}, the square transfer function $G(s)$ is NI if and only if the Hamiltonian matrix, $N_0$ described by (\ref{eq: Hamiltonian_matrix}), has no pure imaginary eigenvalues with odd multiplicity.
\begin{equation}\label{eq: Hamiltonian_matrix}
N_0 = \begin{bmatrix}
    A + BQ_0^{-1}CA & BQ_0^{-1}B^T \\
    -A^TC^TQ_0^{-1}CA & -A^T - A^TC^TQ_0^{-1}B^T
    \end{bmatrix},
\end{equation}
where $$Q_0 = - (CB + B^TC^T).$$
The eigenvalues of the Hamiltonian matrix in (\ref{eq: Hamiltonian_matrix}) for the state-space matrices of the system (\ref{eq:state_space_model}) are obtained as
\begin{align*}
    \begin{bmatrix}
    \pm 1.068\times 10^8\\ \pm 38.45 \\ -0.0043 \pm 41.33i\\ 4.68 \times 10^{-3}\\ \pm 9.56 \times 10^4i
    \end{bmatrix},
\end{align*}
which reveals no pure imaginary eigenvalues, thus the transfer function matrix $G(s)$ with the state space realization of the form (\ref{eq:state_space_model}) is NI. 
It should be noted that the NI property of the nanopositioner can be also investigated using the measured frequency response through the eigenvalues of the matrix $j[G(j\omega) - G^*(j\omega)]$~\cite{mabrok2013spectral}. The NI property of the identified system model obtained from the measured frequency response data implies that the NI property is satisfied over the frequency range of interest. Considering this, we aim to verify whether a HIGS controller can stabilize a TITO MEMS nanopositioner although its dynamics violate the NI property at high frequencies.

\subsection{Controller Design}
According to stability results presented in Section~\ref{sec:MIMO_HIGS}, a stabilizing TITO multi-HIGS is required to satisfy the following conditions
\begin{equation*}
K_h^{-1}-G(0)>0.	
\end{equation*}
Accordingly, $K_h$ is determined by solving a feasibility problem formulated in MATLAB using the YALMIP
toolbox~\cite{Yalmip_2004} and solved with the MOSEK~\cite{Mosek_2019}. Since $\omega_h$ plays no role in stability analysis, it is tuned to achieve a desired performance level. Therefore, we have
\begin{align*}
    K_h &= \begin{bmatrix}
    0.5617  &     0 \\
         0  & 0.6003
    \end{bmatrix},\nonumber \\
    \omega_h &= \begin{bmatrix}
    1.1516 \times 10^4  & 0\\
    0 & 1.1560 \times 10^4 
    \end{bmatrix} \text{(rad/s)}.
\end{align*}

\begin{figure}[h!]
\centering
\psfrag{r}{$r$}
\psfrag{HIGS}{$\mc H$}
\psfrag{y}{$y$}
\psfrag{plant}{$G(s)$}
\includegraphics[width=8.5cm]{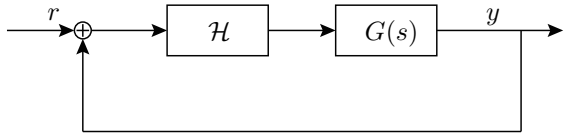}
\caption{Closed-loop interconnection of a TITO multi-HIGS $\mc H$ and the TITO MEMS nanopositioner $G(s)$.}
\label{fig:experiment CL setting}
\end{figure}
The closed-loop interconnection of the TITO multi-HIGS and the TITO MEMS nanopositioner is shown in Fig.~\ref{fig:experiment CL setting}, where $r$ is the reference signal. The closed-loop frequency response of the MEMS nanopositioner in positive feedback with the TITO multi-HIGS is shown in Fig.~\ref{fig:TITO_FRF}. The frequency response is obtained using the frequency response data model of the nanopositioner and the describing function of the multi-HIGS.

\subsection{Experiments}
To assess the controller performance and the stability of the MEMS nanopositioner in a closed-loop interconnection with the multi-HIGS element in TITO configuration, we implement the multi-HIGS controller in real time and perform closed-loop experiments in the time domain. Fig.~\ref{fig:setup_SEM} shows the experimental setup, including the MEMS nanopositioner mounted on a custom-designed PCB with actuation and sensing signal paths, a dSPACE rapid prototyping system, and high-gain voltage amplifiers.

\begin{figure*}[h!]
	\centering
	\includegraphics[width=0.9\linewidth]{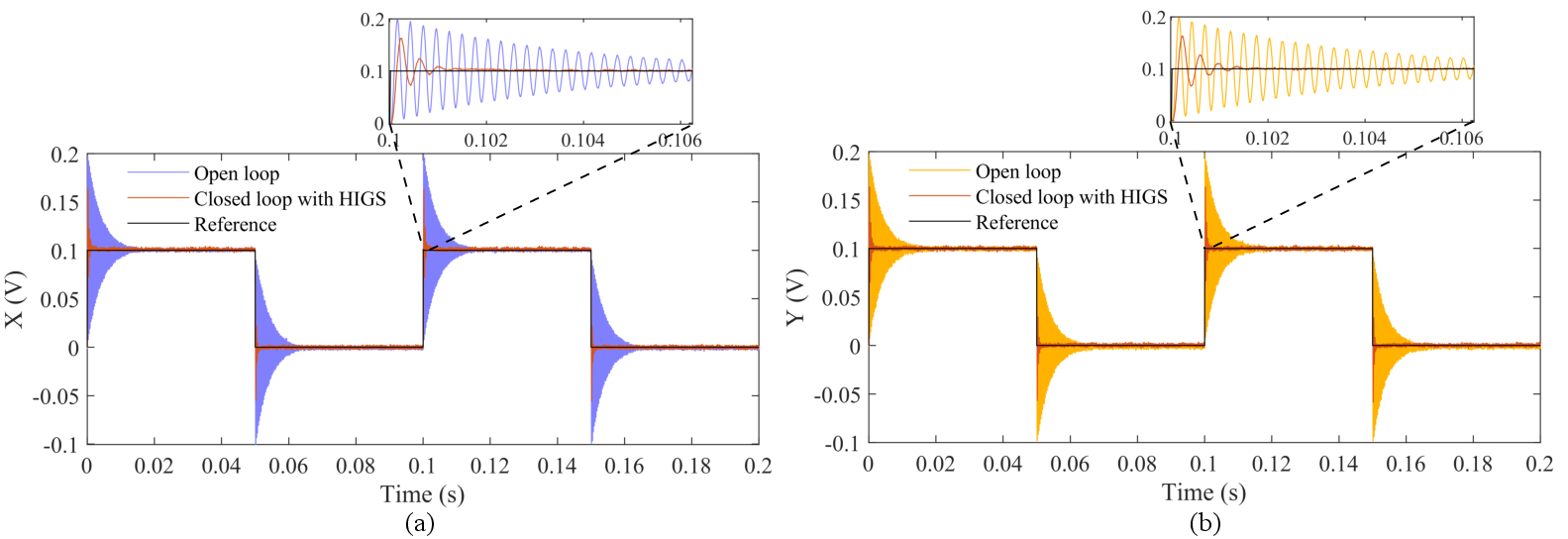}
	\caption{Time-domain response of the MEMS nanopositioner in X and Y axes in open loop and in positive feedback interconnection with the multi-HIGS element in TITO configuration in experiments.}\label{fig:step_response}
\end{figure*}

The multi-HIGS element and the feedback loop were digitally implemented in a dSPACE rapid prototyping system with a sampling rate of $80\, \text{kHz}$. The  X and Y axis sensor outputs were recorded in open loop and closed loop while a sequence of pulses with a frequency of $10\, \text{Hz}$ and amplitude of $0.1\, \text{V}$ was applied to the device as an external disturbance. Fig.~\ref{fig:step_response} shows the nanopositioner sensor outputs in the X and Y axes, respectively. We observe that the closed-loop system with the multi-HIGS is asymptotically stable. This means that a MEMS nanopositioner can be stabilized by a multi-HIGS controller. This experimental result is consistent with our expectations. The reason that the out-of-bandwidth non-NI dynamics do not destabilize the closed-loop system is that at high frequencies, deviations from NI are too insignificant to cause instabilities. This property was observed and explained for the same plant in \cite{Nikooienejad_Data_Driven_2022}. Furthermore, from Fig.~\ref{fig:TITO_FRF}, we observe that the magnitude of the frequency response of the MEMS nanopositioner is bounded below a certain level when the frequency is greater than 3976 Hz. Hence, the stability is achieved via the small-gain theorem; see \cite{patra2011stability,das2014resonant} for the stability of systems with ``mixed" NI and small-gain properties. It can be seen in Fig.~\ref{fig:step_response} that in comparison with the open-loop performance, applying the multi-HIGS in positive feedback improves the performance of the MEMS nanopositioner. From the close-up views, it is clear that fast settling time and reduced overshoot are achieved in both axes. To further improve the overshoot, $\omega_h$ can be further reduced, which in turn limits the closed-loop bandwidth.

\section{CONCLUSION and FUTURE WORK}\label{sec:conclusion}
In this paper, it is shown that a single HIGS, a multi-HIGS, and the cascade of two HIGS are all NNI systems. Using NNI systems theory, these HIGS can be applied as controllers to stabilize linear NI systems. For a linear NI system with a minimal realization, there always exists a HIGS controller such that their positive feedback interconnection is asymptotically stable. We illustrate by a real-world experiment that HIGS can improve the control performance of a MEMS nanopositioner.

To this end, we point out that the results presented in this paper can be extended in a number of directions. One possible future work is to investigate the closed-loop stability when there are external disturbances acting on the system. In this case, instead of asymptotic stability, we would need to determine the input-to-state stability of the system. Another direction is in the construction of a cascaded multi-HIGS, which has a similar parallel structure as the multi-HIGS introduced in Section \ref{sec:MIMO_HIGS}, but with each input-output channel being a cascaded HIGS as introduced in Section \ref{sec:cascaded HIGS}. If such a system can be used to stabilize an NI plant, then it is expected to have advantages different to that of a multi-HIGS.

\section*{ACKNOWLEDGEMENT}
The research performed by Reza Moheimani and Nastaran Nikooienejad was supported by the United States Department of Energy under award number DE-SC0018527. This work was also supported by the Australian Research Council under grant DP190102158.


\section*{APPENDIX}
\noindent\textbf{Proof of Lemma \ref{lemma:F implication}:}

\noindent Consider the following inequality
\begin{equation*}
\left(\sqrt{\frac{1}{k_h}}x_h-\sqrt{k_h}e\right)^2\geq 0,
\end{equation*}
where equality only holds when $x_h=k_he$. Therefore,
\begin{equation*}
\frac{1}{k_h}x_h^2 - 2ex_h + k_he^2 \geq 0.
\end{equation*}
Considering the condition in $\mc F$ as given in (\ref{eq:F}), this implies that
\begin{equation*}
ex_h-k_he^2\leq \frac{1}{k_h}x_h^2 - ex_h \leq 0,	
\end{equation*}
where equality only holds when $x_h=k_he$. \hfill {\small$\blacksquare$}
\\

\noindent\textbf{Proof of Lemma \ref{lemma:parallel NI}:}

\noindent The storage function $\widehat V(X)$ defined in (\ref{eq:sf_parallel_NI}) satisfies
	\begin{equation*}
	\dot {\widehat V}(X) = \Sigma_{i=1}^N \dot V_i(x_i) \leq \Sigma_{i=1}^N u_i^T\dot y_i = U^T\dot Y.
	\end{equation*}
Therefore, the system with input $U$ and output $Y$ also satisfies Definition \ref{def:nonlinear_NI}. \hfill {\small$\blacksquare$}
\\

\noindent \textbf{Proof of Lemma \ref{lemma:MIMO lossless condition}:} 

\noindent Considering (\ref{eq:multi-HIGS NNI ineq}), we have $\dot {\widehat V}_h(X_h)=E_h^T \dot U_h$ only if $\dot V_i(x_{h,i})=e_{i}\dot u_{i}=e_{i}\dot x_{h,i}$ for all $i=1,2,\cdots,N$. For a HIGS of the form (\ref{eq:HIGS}), $\dot V_i(x_{h,i})=e_{i}\dot x_{h,i}$ implies $\frac{1}{k_{h,i}}x_{h,i}\dot x_{h,i}=e_{i}\dot x_{h,i}$. This holds if $\dot x_{h,i}=0$ or $x_{h,i}=k_{h,i}e_{i}$. Consider the condition that $\dot x_{h,i}=0$, in $\mc F_{1,i}$ mode, $\dot x_{h,i}=0$ implies $e_{i}=0$. According to (\ref{eq:Fi}), $e_{i}=0$ implies that $x_{h,i}=0$. In this case, $x_{h,i}=k_{h,i}e_{i}=0$. In $\mc F_{2,i}$ mode, $x_{h,i}=k_{h,i}e_{i}$. Hence, $x_{h,i}=k_{h,i}e_{i}$ always holds in the case that $\dot V_i(x_{h,i})=e_{i}\dot u_{i}=e_{i}\dot x_{h,i}$. Therefore, $\dot {\widehat V}_h(X_h)=E_h^T \dot U_h$ implies that $X_h = K_hE_h$. \hfill {\small$\blacksquare$}
\\

\noindent\textbf{Proof of Lemma \ref{lemma:equality holds condition in F}:}

\noindent We have that $ex_h\equiv\frac{1}{k_h}x_h^2$ implies $x_h\equiv 0$ or $x_h\equiv k_he$. We only consider the case $x_h \equiv 0$ in the $\mc F_1$ mode, because $x_h= k_he$ always holds in the $\mc F_2$ mode. In the $\mc F_1$ mode, $u_h \equiv x_h\equiv 0$ implies that $\dot x_h \equiv 0$. This implies that $e\equiv 0$. Therefore, $x_h\equiv x_h\equiv k_he\equiv 0$. \hfill {\small$\blacksquare$}

\bibliographystyle{IEEEtran}
\end{document}